\documentclass[seceq,supplement]{ptptex}

\usepackage{graphicx}
\usepackage{wrapft}

\title{
Agent-Based Model Approach to Complex Phenomena \\ in Real Economy%
}

\author{
Hiroshi \textsc{IYETOMI}$^{1,}$\footnote{E-mail: hiyetomi@sc.niigata-u.ac.jp},
Hideaki \textsc{AOYAMA}$^{2}$,
Yoshi \textsc{FUJIWARA}$^{3}$,\\
Yuichi \textsc{IKEDA}$^{4}$ and
Wataru \textsc{SOUMA}$^{3}$%
}

\inst{
$^1$Department of Physics, Niigata University, Niigata 950-2181, Japan
\\
$^2$Physics Department, Kyoto University, Kyoto 606-8501, Japan
\\ 
$^3$NiCT/ATR CIS, Applied Network Science Lab., Kyoto 619-0288, Japan
\\
$^4$Hitachi Research Institute, Tokyo 101-8010, Japan
}

\abst{
An agent-based model for firms' dynamics is developed. The model consists of firm agents with identical characteristic parameters and a bank agent. Dynamics of those agents is described by their balance sheets. Each firm tries to maximize its expected profit with possible risks in market. Infinite growth of a firm directed by the ``profit maximization" principle is suppressed by a concept of ``going concern". Possibility of bankruptcy of firms is also introduced by incorporating a retardation effect of information on firms' decision. The firms, mutually interacting through the monopolistic bank, become heterogeneous in the course of temporal evolution. Statistical properties of firms' dynamics obtained by simulations based on the model are discussed in light of observations in the real economy. 
}

\begin{document}

\maketitle

\section{Introduction}
\label{sec:1}

An agent-based model is used to elucidate complex phenomena encountered in a wide variety of social and economic systems.\cite{Schweitzer2003} \ The agent is a natural extension of the atomic concept worked out for describing physical systems. Agents have internal structures characterized by different parameters. Agents are made so intelligent as to be autonomous; for instance, an agent has a capability to adapt itself to surrounding conditions. Also agents interact with each other according to simple rules. A complex system is thus regarded as an assembly of interacting agents.

It is a dream for physicists to explain complex phenomena happening in the economic
world based on a simple model. A promising model is that consisting of interacting agents.\cite{Aoki2002, AY2007, DGGGP2008} \ Gallegati and his collaborators~\cite{DGGGP2008, GGK2003, Fujiwara2004} constructed a workable model, and showed that it successfully reproduced a set of stylized facts including the distribution with a power-law tail of firms' size and a Laplace-type distribution of the growth rate of firms. 

A large collection of firm agents with identical characteristic parameters and a monopolistic bank constitute the model. Dynamics of the agents are characterized by their balance sheets. Each firm tries to maximize its expected profit with possible risks in market. The firms, mutually interacting through the bank, become heterogeneous in the course of temporal evolution. Possibility of bankruptcy of a firm is also taken into account. Such a microscopic model, once established, enables us to investigate interplay between behavior of individual firms and macroscopic trend of economy. We are now in a stage to be capable of calibrating the model thanks to accumulation of results for statistical properties of dynamics of real firms.\cite{Stanley1996, Takayasu1999, Axtell2001, FGAGS2004, Ishikawa2005}.

The objective of the present study is to develop an agent-based model for dynamics of firms along the line laid by Gallegati \textit{et al.}\cite{DGGGP2008, GGK2003, Fujiwara2004} \ We reconstruct the model to elucidate the conceptual ingredients. Compromise between the two concepts, ``profit maximization" and ``going concern" plays a key role along with decision of firms using imperfect information on their financial conditions. Simulations based on the model are then carried out for statistical properties of firms' dynamics and the results so obtained are discussed in light of observations in the real economy. A preliminary account of the present work has been given in Ref.~\citen{Iyetomi2005}.

\section{Agent-Based Modeling}

The agent-based modeling of firms' dynamics due to Gallegati \textit{et al.} provides us with a sound starting point. The system consists of a swarm of firm agents with identical characteristic parameters and a monopolistic bank agent, as depicted in Fig.~\ref{fig:ABM}. The firms interact to each other only through the bank; direct interactions between firms due to business network are not taken into account. We first modify the model to make it most rational within the original framework; each firm draws up its production plan with all information available at hand. And then we introduce retardation of information on firms' dynamics. To present our model we will follow the same notations and the same values for the model parameters in Ref.~\citen{GGK2003} wherever it is possible. 

\subsection{Balance Sheet Dynamics}
\begin{figure}
\centering
\includegraphics[height=5cm,bb=22 248 577 643]{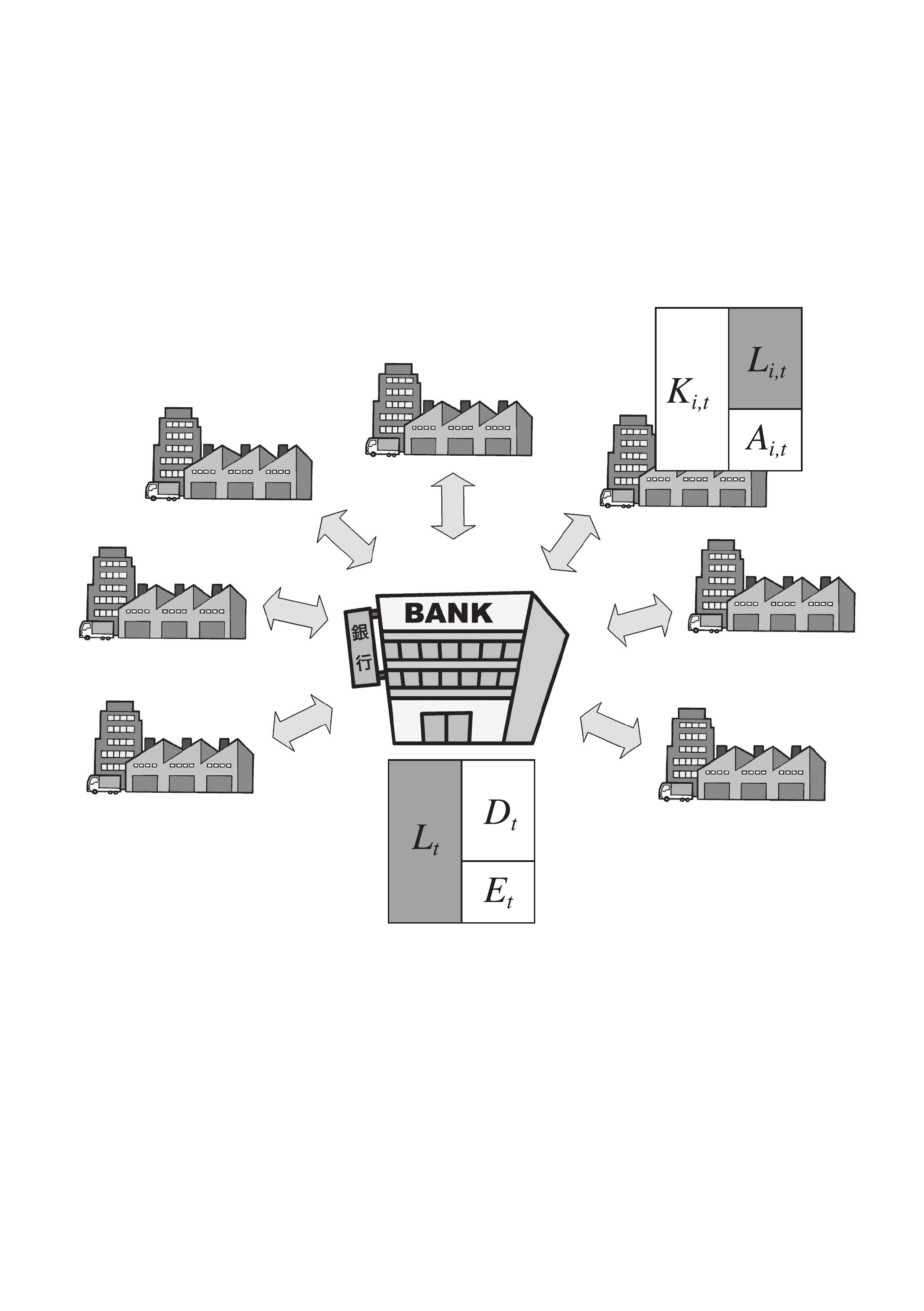}
\caption{Agent-based model consisting of multiagents of firms and a single agent of bank.}
\label{fig:ABM}       
\end{figure}

Dynamics of the agents are described in terms of balance sheets (see Fig.~\ref{fig:ABM}). A firm agent $i$ has total assets $K_{i,t}$ and liabilities $L_{i,t}$ from the bank at the beginning of a time period $t$. According to the accounting equation, equity capital $A_{i,t}$ of the firm must equal assets minus liabilities. On the other hand, the bank agent has a balance sheet on which its aggregate supply of credit, $L_{t}=\sum_{i} L_{i,t}$, is balanced by sum of total deposits $D_{t}$ and equity capital $E_{t}$. Stocks and flows are two kinds of basic variables to construct system dynamics models. The balance sheets have only stock variables so that they are just snapshots of financial conditions of the agents in time. Flow variables such as profit and investment determine evolution of the economic system. 

\begin{figure}
\centering
\includegraphics[height=6cm,bb=0 150 595 690]{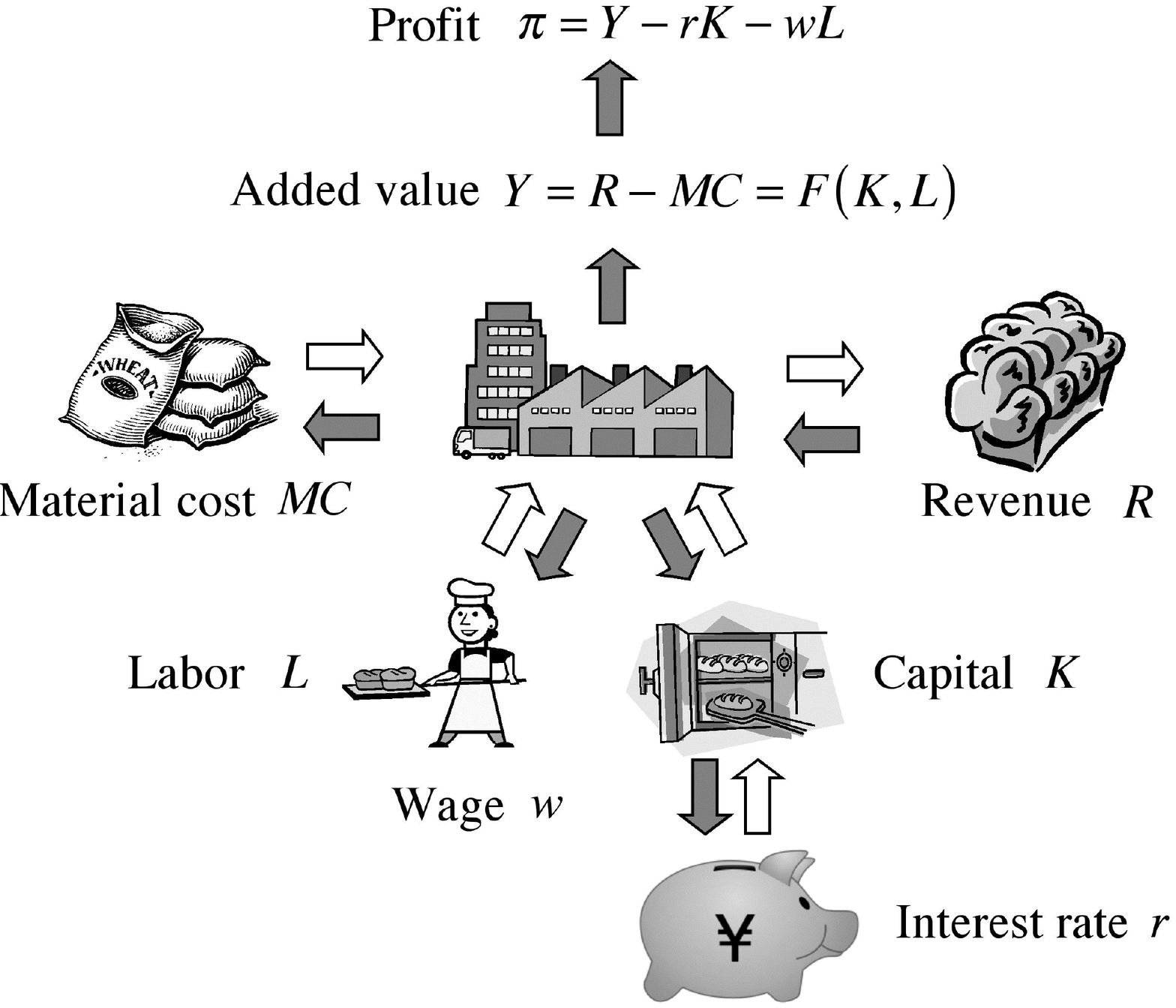}
\caption{Production activity of a firm.}
\label{fig:firm_activity}       
\end{figure}

\subsection{Firm Agent}

\begin{figure}
\centering
\includegraphics[width=6cm,bb=96 222 498 611]{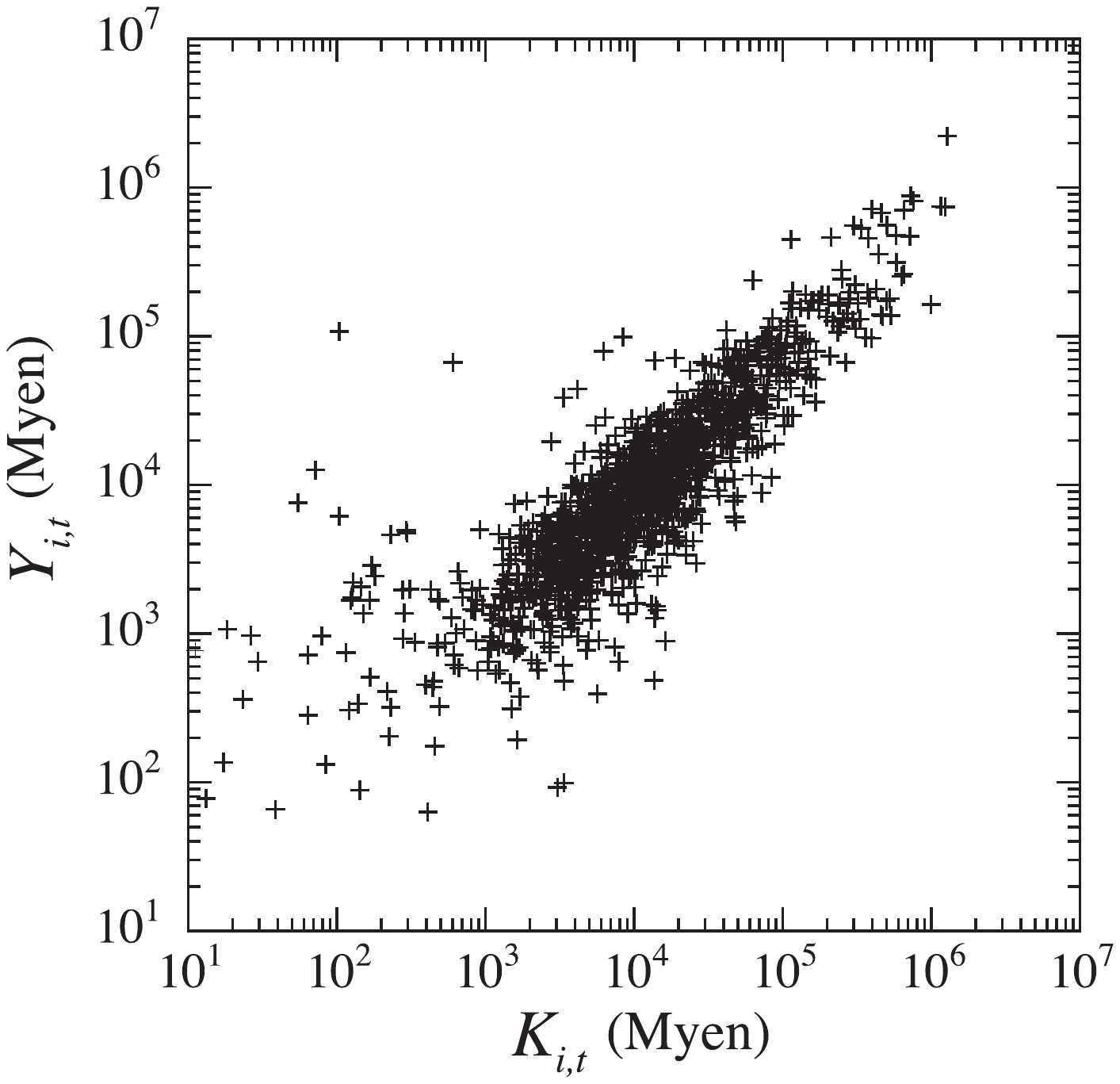}
\caption{Added value $Y_{i,t}$ versus fixed assets $K_{i,t}$ for the listed Japanese firms in 2006.}
\label{fig:KvsY_NEEDS2006makers}
\end{figure}

Figure~\ref{fig:firm_activity} depicts the basic activity of a firm; it features a bakery. The firm purchases materials to make products and sells them to obtain money. Subtraction of material cost $MC$ from the revenue $R$ gives added value $Y$ for the firm. The firm inevitably needs plants and employees for its production activity. To calculate profit, therefore, we further have to subtract financial and labor costs necessary to keep the capital $K$ and the labor $L$. One of fundamental ideas in the economic theory is that $Y$ is a function of input variables, $K$ and $L$: 
\begin{equation}
\label{pf}
Y=F\left( K,L \right).
\end{equation}
For simplicity in modeling firms' dynamics, we concentrate on the financial aspect of the production function (\ref{pf}) and assume the added value linearly scales to the capital input:
\begin{equation}
\label{pf1}
Y=\phi K,
\end{equation}
where the proportionate $\phi$ is taken as $\phi=0.1$. Figure~\ref{fig:KvsY_NEEDS2006makers} validates this modeling for the production function.

At the beginning of a given time period $t$ the $i$-th firm changes its total asset $K_{i,t}$ to maximize the expected value of profit. This strategic behavior of the firm, called ``profit maximization", is a well-known hypothesis in economics since Adam Smith, although it has not been confirmed yet.

The profit of a firm is fixed at the end of each period such as
\begin{equation}
\label{pi}
\pi_{i,t}=u_{i,t}Y_{i,t}-r_{i,t}K_{i,t}=\left( u_{i,t}\phi-r_{i,t}\right) K_{i,t},
\end{equation}
where $r_{i,t}$ is an interest rate for the financial cost. The parameter $u_{i,t}$ reflects uncertainty in a market. Since a market consists of a huge number of economic degrees of freedom, determination of the selling price becomes inevitably stochastic. We also assume that $u_{i,t}$ is independent of the firm size in harmony with Gibrat's law. We thus take $u_{i,t}$ as a uniform random number in (0,2); this is an arbitrary choice as adopted in the original model.

If a firm takes an aggressive production plan, it has a finite probability of bankruptcy. Bankruptcy of a firm is defined at the end of a period $t$ by the condition,
\begin{equation}
\label{A}
A_{i,t}=A_{i,t-1}+\pi_{i,t-1} < 0.
\end{equation}
Substitution of (\ref{pi}) into (\ref{A}) results in the following formula for the bankruptcy probability: 
\begin{equation}
\label{PB}
P_{\mathrm{B}} \left( {K_{i,t} } \right) = \begin{cases}
     {\frac{{r_{i,t} K_{i,t}  - A_{i,t} }}{{2\phi K_{i,t} }}} &
                 \text{for ${K_{i,t} {\rm{ > }}\frac{{A_{i,t} }}{{r_{i,t} }}}$}, \\
     0 & \text{otherwise}.
\end{cases}
\end{equation}
We thus see there is a upper bound in the size for a firm to be free from bankruptcy.

Another management policy, called ``going concern", prevents a firm from expanding its size infinitely; a firm desires to survive forever. We assume here that firms adopt a solid production plan with a safety factor $\sigma (\leq 1)$:
\begin{equation}
\label{K}
K_{i,t}=\sigma\frac{A_{i,t}}{r_{i,t}}.
\end{equation}
This choice compromises the two directly-opposed economic ideas. 

\begin{figure}
\centering
\includegraphics[height=5cm,bb=98 321 467 638]{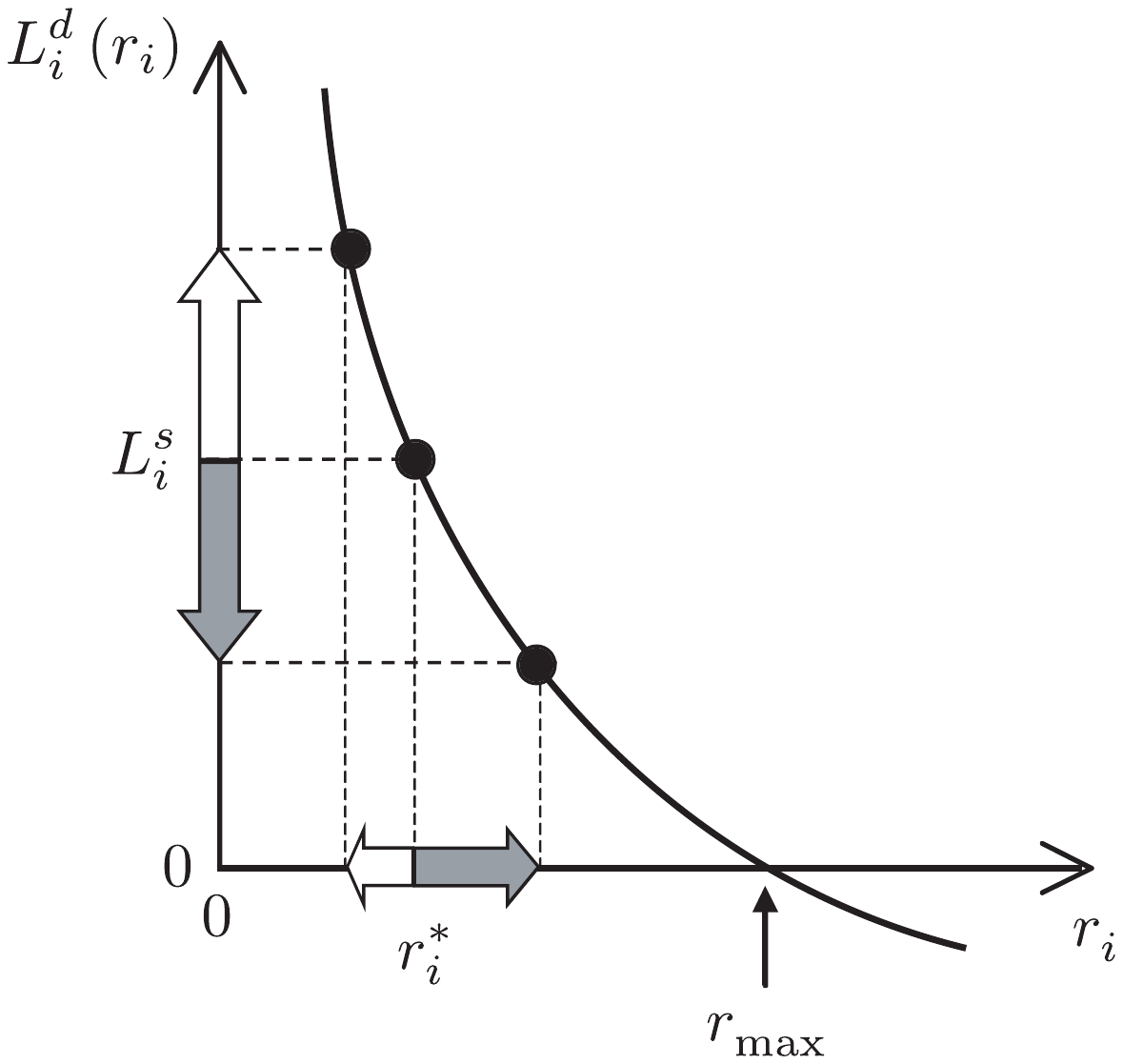}
\caption{Determination of the interest rate for the financial cost of a firm.}
\label{fig:interest}       
\end{figure}

The interest rate for each firm is then determined through demand and supply balance in credit market between firms and the bank. The firm requests the bank to finance the following amount of money derived from (\ref{K}):
\begin{equation}
\label{L_demand}
L_{i,t}^{d}=K_{i,t}-A_{i,t}=\left( \frac{\sigma}{r_{i,t}}-1 \right) A_{i,t}.
\end{equation}
On the other hand, credit is granted to the firm by the bank in proportion to its relative size in the preceding period as
\begin{equation}
\label{L_supply}
L_{i,t}^{s}=L_t \frac{K_{i,t-1}}{\sum\limits_i {K_{i,t-1}}},
\end{equation}
Balancing (\ref{L_demand}) and (\ref{L_supply}) gives the formula for the interest rate,
\begin{equation}
\label{interest}
r_{i,t}=\frac{\sigma A_{i,t}}{L_{i,t}^{s}+A_{i,t}}.
\end{equation}
Such an equilibrium mechanism to determine the interest rate is depicted in Fig.~\ref{fig:interest}. The maximum rate is given by $r_\mathrm{max} = \sigma$. If the firm obtain more credit from the bank, the interest rate decreases, and vice versa.

\subsection{Bankruptcy}

Firm agents with the behavioral rules above-mentioned show no bankruptcy. Real firms, however, are always afraid of being bankrupted. To incorporate a possibility of bankruptcy for firms into the model, we replace the equity capital of the current period by that of the preceding period in (\ref{K}):
\begin{equation}
\label{K1}
K_{i,t}=\sigma\frac{A_{i,t-1}}{r_{i,t}}.
\end{equation}
Firms thus determine their production plans with delayed information. This replacement turns over the conservative attitude of firms when they are in a recession phase. The firms incidentally take speculative management actions.
We arbitrarily set $\sigma = 1/2$, which enables us to make a smooth connection with the original formulae. For instance, the equation (\ref{K1}) is deduced by omitting the intensive terms, independent of the firm size, in the corresponding equation in Ref.~\citen{GGK2003}. 

The delay of information is one of causes of bankruptcy for firms. Alternatives include existence of unexpected risk and propagation of bankruptcy akin to a chain reaction. Here the possible risk is supposed to be totally predictable by specifying a definite range for $u_{i,t}$ in (\ref{pi}) , but nobody can avoid unexpected risk in real business. In fact, firms are linked to each other through transaction with supply of credit. If a large firm is bankrupted, then a credit risk shock will propagate over the network. The chain reaction bankruptcy arising from direct interactions among firms is out of scope in the present study.

\subsection{Bank Agent}

We assume that 
the bank expands its business subject to minimum requirement of a prudential rule with a risk coefficient $\alpha$:
\begin{equation}
\label{bank_L}
L_t  = \sum\limits_i {L_{i,t}}  = \frac{{E_t}}{\alpha}.
\end{equation}
The Basel committee of the Bank for International Settlements introduced an international capital adequacy standard called Basel I in 1988.\footnote{The Basel I is to be replaced by the Basel II with more refined rules.} It requires that each bank has its capital being at least 8\% of the total asset. So we take here $\alpha = 0.08$.

The bank derives a profit through investing its money in firms. However its net profit $\Pi _t$ is given by subtracting financial costs from the sum of interests:  
\begin{equation}
\label{bank_pi}
\Pi _t  = \sum\limits_i {r_{i,t} L_{i,t} }  - r_t \left( {1 - \omega } \right)D_t  - r_t E_t  - \sum\limits_i {^{'}B_{i,t}},
\end{equation}
where
\begin{equation}
r_t = \frac{1}{N} \sum\limits_i {r_{i,t} }.
\end{equation}
The second and third terms in the right-hand side of (\ref{bank_pi}) stand for interests paid to depositors and to investors, respectively; the profit margin $\omega$ is set as $\omega = 0.002$. The last term takes account of additional loss due to bad debts stemming from bankruptcy of firms:

\begin{equation}
\label{E_bank}
E_{t}=E_{t-1}+\Pi_{t-1}.
\end{equation}

\subsection{Comparison with the Previous Model}
Bankruptcy costs with quadratic dependence on added value were introduced in the original modeling due to Gallegati \textit{et al.}\cite{DGGGP2008, GGK2003, Fujiwara2004} \ It was necessary to prevent firms from growing unlimitedly. Here such a rather obscure idea is replaced by the transparent principle of ``going concern''.
Also the original model suffers from the followings: i) financial quantities at the beginning of a period are not clearly distinguished from those at the end of the period, ii) the bankruptcy probability of a firm is not prohibited to take a negative value, iii) extensive and intensive quantities are mixed up in the formulae. The present modification of the model cures these problems.

\section{Representative Agent}
\begin{figure}
  \centering
  \includegraphics[height=5.5cm,bb=105 122 778 435]{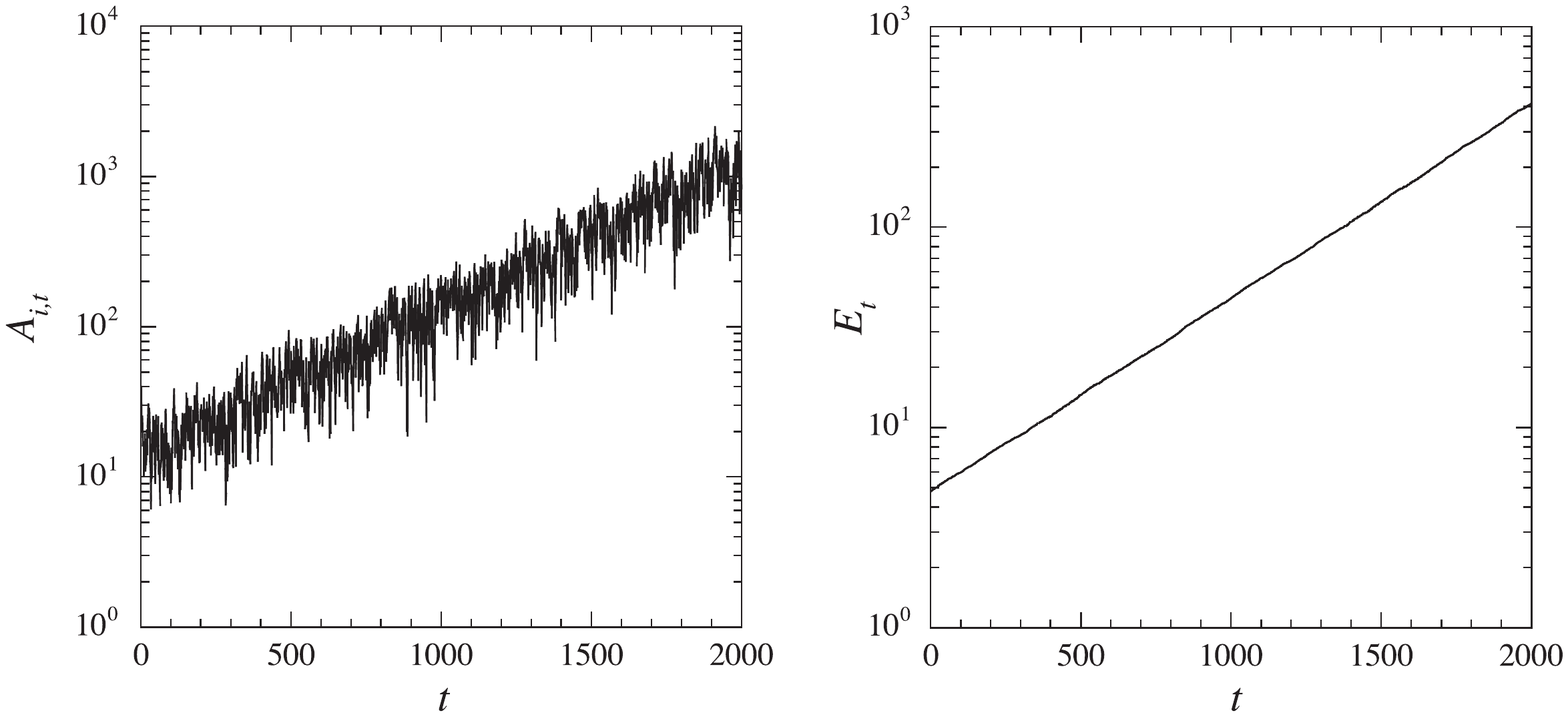}
  \caption{Numerical results for the representative agent model. The left and right panels show temporal change of the equity capital of the representative firm and that of the bank agent, respectively.}
  \label{fig:rep_agent}
\end{figure}

We first study a system comprising a single ideal firm interacting with the bank agent; firms are thus represented by the single agent. This representative agent model, neglecting heterogeneity of agents, is a traditional approach in economics. Figure~\ref{fig:rep_agent} shows that both agents grow exponentially. This is an intrinsic property of the present model as seen below. 

For a market with no fluctuations in the selling price ($u_{i,t}=1$) and in the interest rate ($r_{i,t}=r$), one can obtain an analytic solution for each agent. The evolutionary equations for the ideal firm and bank agents now read 

\begin{equation}
\label{firm_analytic}
\overline {A_{t + 1} }  = \overline {A_t }  + \left( {\phi  - r} \right)\overline {K_t }  = \left( {1 + \sigma \frac{{\phi  - r}}{r}} \right)\overline {A_t },
\end{equation}

\begin{equation}
\label{bank_analytic}
\overline {E_{t + 1} }  = \overline {E_t }  + \omega r\left( {\overline {L_t }  - \overline {K_t } } \right) = \left( {1 - \omega r + \frac{{\omega r}}{\alpha }} \right)\overline {E_t }.
\end{equation}
Then analytic solutions to (\ref{firm_analytic}) and (\ref{bank_analytic}) are obtained as
\begin{equation}
\overline {A_{t}}  \propto \exp \left[ {t\ln \left( {1 + \sigma \frac{{\phi  - r}}{r}} \right)} \right],
\end{equation}
\begin{equation}
\overline {E_{t}}  \propto \exp \left[ {t\ln \left( {1 - \omega r + \frac{{\omega r}}{\alpha }} \right)} \right].
\end{equation}
Equating the two formulae for the growth rate, one can determine $r$ as
\begin{equation}
\label{r_analytic}
r = \frac{{ - 1 + \sqrt {1 + 4\phi \xi } }}{{2\xi }} \simeq \phi \left( {1 - \phi \xi } \right) = 0.0995,
\end{equation}
where
\begin{equation}
\xi  = 2\omega \left( {\frac{1}{\alpha } - 1} \right) = 0.046 \ll 1.
\end{equation}
The growth rate derived from (\ref{r_analytic}) explains the results in Fig.~\ref{fig:rep_agent} very well.

\begin{figure}
  \centering
  \includegraphics[height=5cm,bb=28 156 786 427]{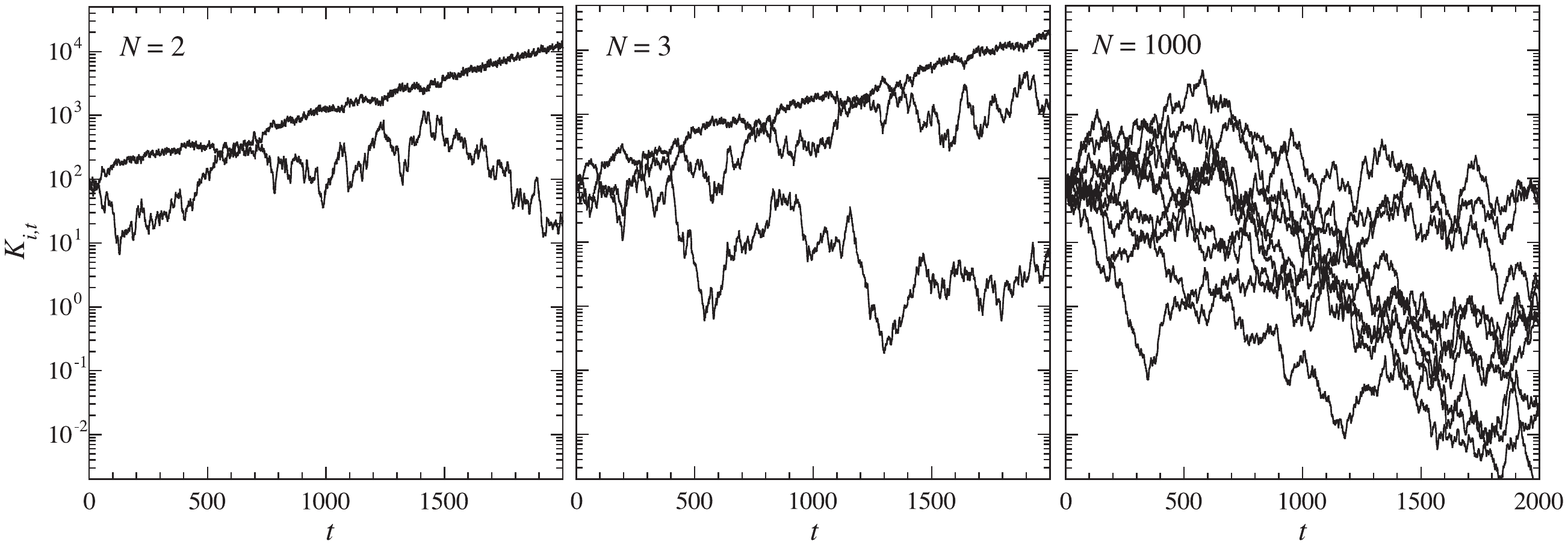}
  \caption{Temporal change of the total assets of firms in the system with $N=2,3,1000$. For $N=1000$, ten firms were selected arbitrarily.}
  \label{fig:sim_N_dep}
\end{figure}

Figure~\ref{fig:sim_N_dep} shows results obtained by simulations with $N=2,3,1000$. We see that heterogeneity of firms spontaneously arises from competition among firms interacting through the bank.

\section{Simulations with Multiagents}

We executed numerical simulations in the present model with the same characteristic parameters and initial conditions for the agents as those in Ref.~\citen{GGK2003}, but with a much larger number of firms ($N=100,000$).

\subsection{Growing Economy}
\begin{figure}
  \centering
  \includegraphics[height=5.5cm,bb=75 119 793 450]{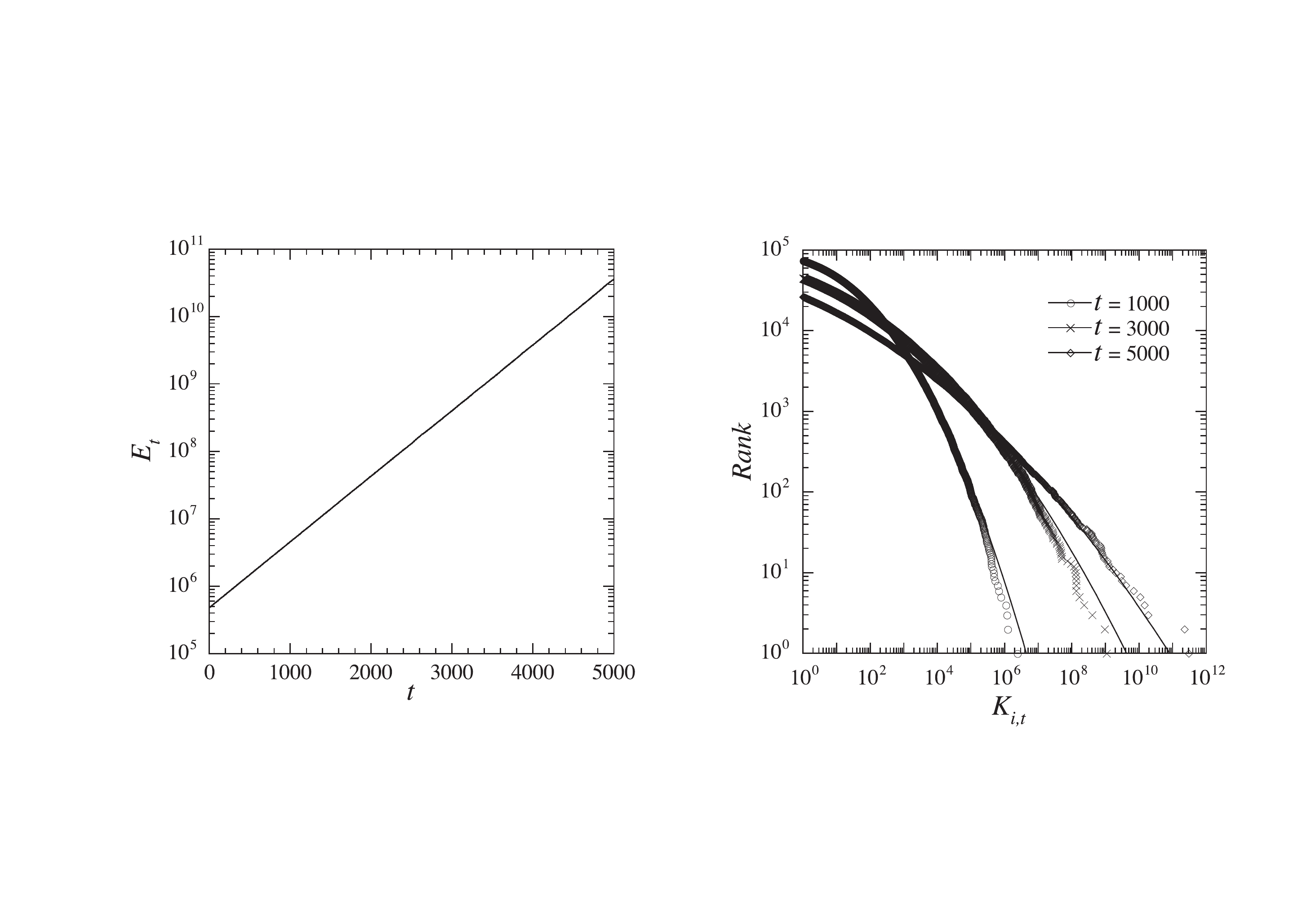}
  \caption{Results of a simulation for firms with perfect information. The left panel shows temporal change of the equity capital of the bank; the right panel, the cumulative distributions of the size of firms at three periods with each fitted to the log-normal distribution (solid curve).}
  \label{fig:firms_perfect}
\end{figure}

\begin{figure}
  \centering
  \includegraphics[height=5.5cm,bb=59 129 771 460]{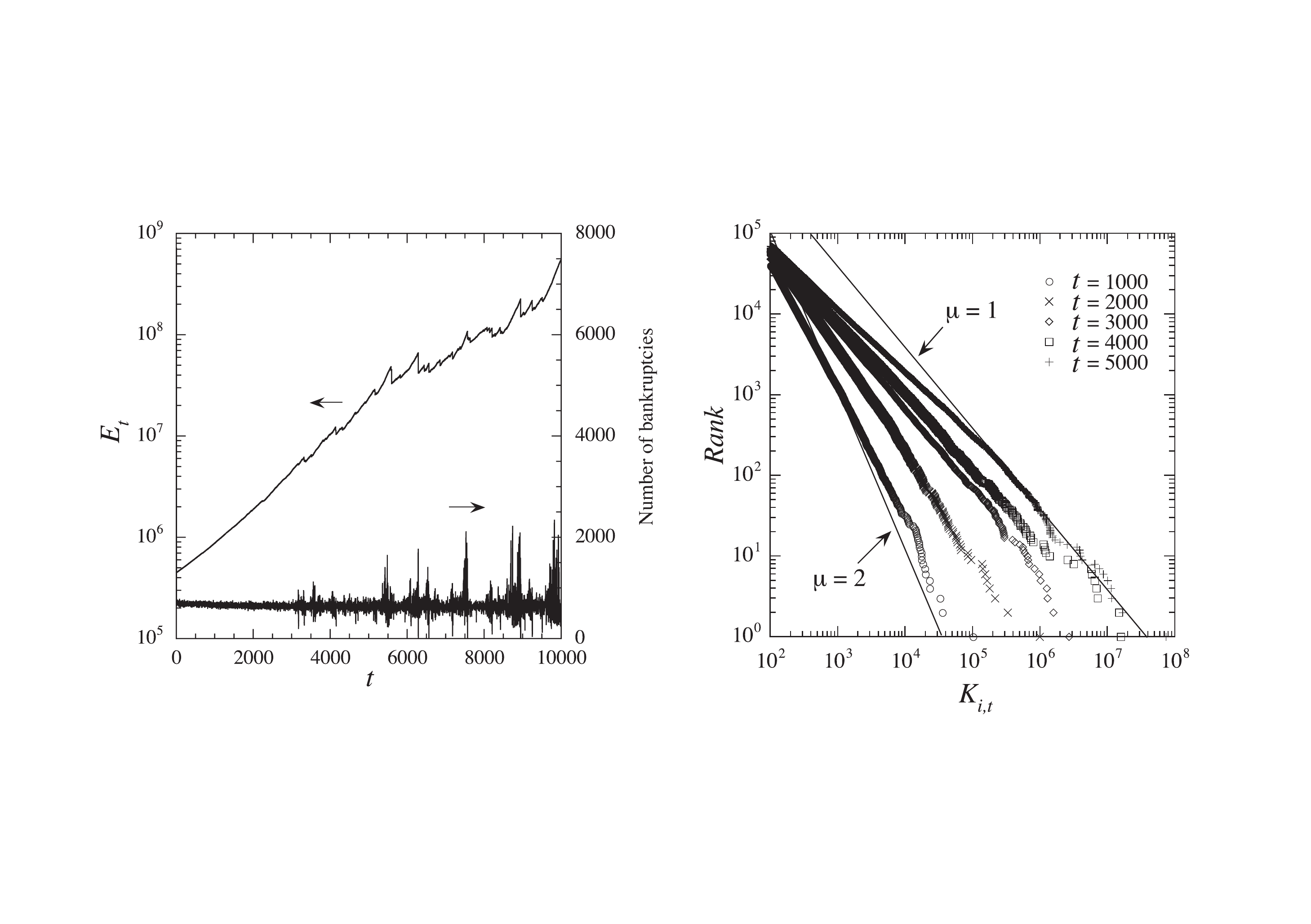}
  \caption{Same as Fig.~\ref{fig:firms_perfect}, but for firms with imperfect information.}
  \label{fig:firms_imperfect}
\end{figure}

As shown in the left panel of Fig.~\ref{fig:firms_perfect}, the growth of the bank is very steady for the rational firms; the growth rate is indistinguishable from that for the representative firm. The corresponding panel of Fig.~\ref{fig:firms_imperfect} shows replacement of the rational firms by the irrational ones gives rise to bankruptcy of firms and hence idiosyncratic shocks in the bank dynamics. Heterogeneity of firms in their sizes simultaneously emerges as demonstrated in the right panels of Figs.~\ref{fig:firms_perfect} and \ref{fig:firms_imperfect}.
The size distribution of the rational firms is well fitted to the log-normal distribution. In contrast, the sizes of the irrational firms are asymptotically distributed in a power-law form,
\begin{equation}
\label{eq:rank}
Rank \propto K_{i,t}^{-\mu}.
\end{equation}
The exponent varies from $\mu \simeq 2$ to $\mu \simeq 1$ as time proceeds. Also bankruptcy of firms happens not uniformly in time.

We thus see the size distribution of firms and the temporal evolution of the bank critically depend on whether firms make full use of available information on their financial conditions or not in determining a production plan for the next period. 

\subsection{Stationary Economy}
\begin{figure}
\centering
\includegraphics[height=5.5cm,bb=72 106 784 437]{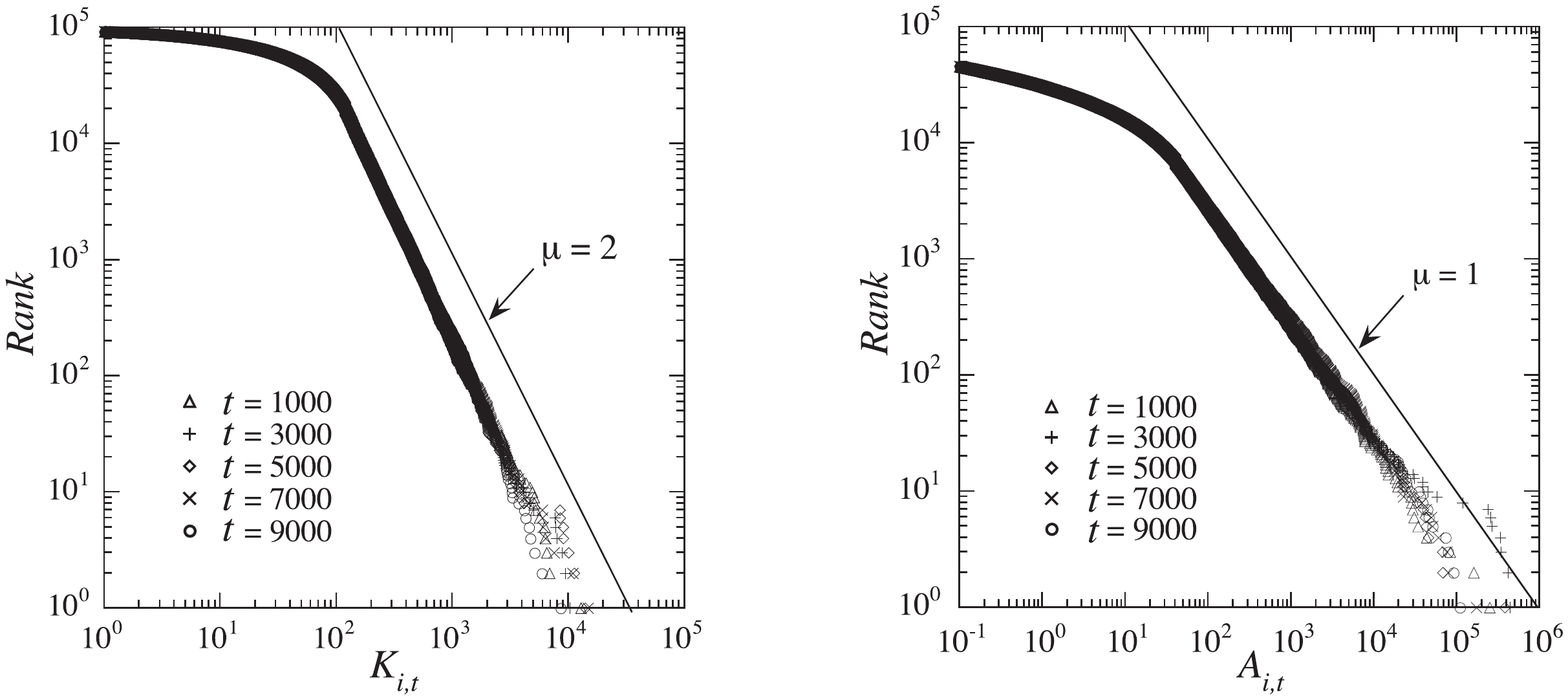}
\caption{Size distribution of actual firm agents in a stationary economy with (left panel) and without (right panel) the interaction with the bank.}
\label{fig:stationary_economy}
\end{figure}

If we control the macroscopic economy by keeping the size of the bank fixed, we have a stationary state for the irrational firms. The distribution of firms' size is a power law  with $\mu \simeq 2$ as shown in the left panel of Fig.~\ref{fig:stationary_economy}. Comparison of it with the right panel of Fig.~\ref{fig:firms_imperfect} reveals an intimate relationship between behavior of individual firms and macroscopic trend of economy; this manifests a micro-macro loop. 

One can terminate the interaction by assuming
\begin{equation}
r_{i,t} = \phi.
\label{eq:cond_rgm}
\end{equation}
This assumption reduces the present agent-based model to a random growth model:
\begin{equation}
A_{i,t+1} = A_{i,t}+\frac{u_{i,t}-1}{2} A_{i,t-1}.
\label{eq:rgm}
\end{equation}
The profit for firms vanishes on average in this situation. The simulation based on (\ref{eq:rgm}) certainly results in a stationary distribution of firms' size with a power-law tail as shown in the right panel of Fig. \ref{fig:stationary_economy}.\footnote{Note that $A_{i,t}$ is simply proportional to $K_{i,t}$ under the condition (\ref{eq:cond_rgm}).}\  But the exponent close to unity is totally different from that for the interacting firms. We thus see that the interactions among firms through the bank give rise to profound changes in the statistical properties of firms as manifested by the variation of the power-law exponent.

\subsection{Synchronized Bankruptcy} 
\begin{figure}
  \centering
  \includegraphics[height=5.5cm,bb=118 404 629 818]{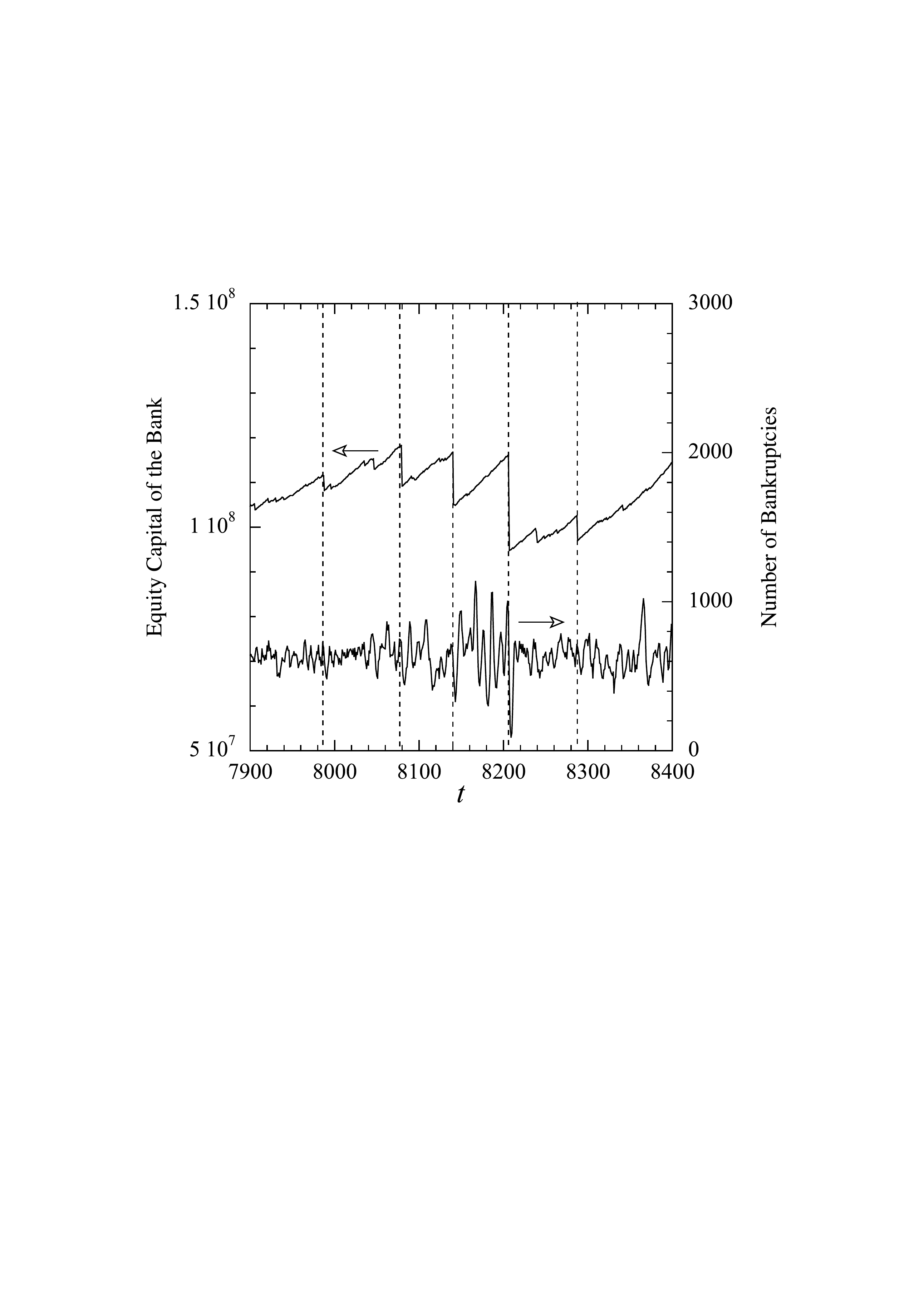}
  \caption{Bankruptcies of firms synchronized with macroscopic shocks reflected in the equity capital of the bank.}
  \label{fig:bankruptcy}
\end{figure}

Figure~\ref{fig:bankruptcy} demonstrates bankruptcies of firms take place in a synchronized way with macroscopic shocks reflected in the equity capital of the bank. Bankruptcy of a large firm triggers such a chain reaction of bankruptcy of firms in the present model which takes into account interactions among firms. The large bankruptcy first gives rise to large bad debt for the bank. Then the equity capital of the bank shrinks and accordingly money supply to firms by the bank decreases. This leads to increase of interest rates for loans from the bank and hence decrease of profits for firms. Financially fragile firms with a low equity ratio $A_{i,t}/K_{i,t}$ are thus strongly influenced by the bankruptcy of the large firm.

\section{Concluding Remarks}
We have clarified the economic ideas built in the model by Gallegati \textit{et al.} Here firms harmonize the ``profit maximization" principle with the ``going concern" concept when making their production plans. The retardation of information in firms' decision-making leads to bankruptcy of firms. Successful results of the original model are carried over to the present model. 

In fact, firms form a complex network through their business transactions. A firm obtains materials from suppliers (upstream firms) and sells its products to customers (downstream firms). Recent studies clearly show the business network has the scale-free nature.\cite{{bib:FAS2006},{bib:FA2008}} \ One can easily guess that some firms have a number of connections acting as hubs. More precisely, it is characterized by a power-law distribution of the number of links (degree) originating from a given node. Why and how such a fascinating network has been formed through the daily activities of firms? To make the present agent-based model for firms' dynamics more predictable, 
it is thus inevitable to take an explicit account of direct interactions between firms over the business network. Addressing to these issues is one of possible ways to proceed for econophysics.\cite{BDGGGS2007, WB2007}

\section*{Acknowledgements}
We thank Taisei Kaijozi for his participation at the early stage in carrying out the present work. We also appreciate the Yukawa Institute for Theoretical Physics at Kyoto University. Discussions during the YITP workshop YITP-W-07-16 on ``Econophysics III: Physical Approach to Social and Economic Phenomena" were useful to complete this paper.

%

\end{document}